\documentclass[twocolumn,,showpacs,preprintnumbers,amsmath,amssymb]{revtex4}
\usepackage{color}
\usepackage{graphicx}% Include figure files
\usepackage{dcolumn}% Align table columns on decimal point
\usepackage{bm}

\begin{document}
\title{Suppression of spin-density-wave transition and emergence of ferromagnetic ordering of Eu$^{2+}$ moments in
EuFe$_{2-x}$Ni$_{x}$As$_{2}$}
\author{Zhi Ren, Xiao Lin, Qian Tao, Shuai Jiang, Zengwei Zhu, Cao Wang, Guanghan Cao\footnote[1]{Electronic address: ghcao@zju.edu.cn} and Zhu'an Xu\footnote[2]{Electronic address: zhuan@zju.edu.cn}}
\affiliation{$^{1}$Department of Physics, Zhejiang University,
Hangzhou 310027, China}

\begin{abstract}
We present a systematic study on the physical properties of
EuFe$_{2-x}$Ni$_{x}$As$_{2}$ (0$\leq$\emph{x}$\leq$0.2) by
electrical resistivity, magnetic susceptibility and thermopower
measurements. The undoped compound EuFe$_{2}$As$_{2}$ undergoes a
spin-density-wave (SDW) transition associated with Fe moments at 195
K, followed by antiferromagnetic (AFM) ordering of Eu$^{2+}$ moments
at 20 K. Ni doping at the Fe site simultaneously suppresses the SDW
transition and AFM ordering of Eu$^{2+}$ moments. For $x\geq$0.06,
the magnetic ordering of Eu$^{2+}$ moments evolves from
antiferromagnetic to ferromagnetic (FM). The SDW transition is
completely suppressed for $x\geq$0.16, however, no superconducting
transition was observed down to 2 K. The possible origins of the
AFM-to-FM transition and the absence of superconductivity in
EuFe$_{2-x}$Ni$_{x}$As$_{2}$ system are discussed.

\end{abstract}


\maketitle \maketitle
\section{\label{sec:level1}INTRODUCTION}
The discovery of superconductivity up to 56 K in iron-based
arsenides \cite{Hosono,Chen-Sm,WNL-Ce,Ren-Pr,Ren-Nd,WHH,Wang-Th} has
aroused great interest in the community of condensed matter physics.
The undoped parent compounds adopt the tetragonal structure at room
temperature, which consists of [Fe$_2$As$_2$]$^{2-}$ layers
separated alternatively by [Ln$_{2}$O$_{2}$]$^{2+}$
\cite{Johnson&Jeitschko,Quebe} or $A^{2+}$ ($A$=Ca, Sr, Ba, Eu)
layers\cite{Pfisterer1980,Pfisterer1983,EuFeAs,CaFeAs ChenXH} . At
low temperatures, the parent compounds undergo a structural phase
transition from tetragonal to orthorhombic,
accompanied\cite{BaFe2As2} or followed\cite{DaiPC Neutron} by a
SDW-like antiferromagnetic (AFM) phase transition. Doping with
electrons or holes in the parent compounds suppresses the phase
transitions and induces the high temperature superconductivity. This
intimate connection between superconductivity and magnetism suggests
unconventional superconductivity in the iron-based
arsenides.\cite{Kotliar,Cao,Singh}

Very recently, superconductivity has been observed in
LaFe$_{1-x}$\emph{M}$_{x}$AsO \cite{Co-doping Sefat,Co-doping
Cao,Ni-doping Cao} and
BaFe$_{2-x}$\emph{M}$_{x}$As$_{2}$\cite{Co-doping Ba,Ni-doping Ba}
($M$=Co and Ni). These findings are quite remarkable and challenge
our common wisdom of superconductivity, which shows that direct
doping in the superconducting-active blocks generally destroys
superconductivity. In high \emph{T}$_{c}$ cuprates, actually, Ni
substitution for Cu in the CuO$_{2}$ planes drastically reduces
\emph{T}$_{c}$. Hence these experimental results provide clues to
the superconducting mechanism for the iron-based arsenide
superconductors. Currently, an itinerant scenario within rigid band
model is more favored to understand this unusual doping-induced
superconductivity.\cite{Co-doping SrFe2As2}

EuFe$_2$As$_2$ is a unique member in the ternary iron arsenide
family due to the fact that Eu$^{2+}$ ions carry local moments,
which orders antiferromagnetically below 20 K.\cite{EuFeAs,EuFeAs
Ren,EuFeAs jeevan} Except this AFM transition, the physical
properties of EuFe$_{2}$As$_{2}$ were found to be quite similar with
those of its isostructural compounds BaFe$_{2}$As$_{2}$ and
SrFe$_{2}$As$_{2}$,\cite{EuFeAs Ren} both of which become
superconducting upon appropriate doping\cite{BaK rotter,SrK wang,SrK
zhu}. It was then expected that EuFe$_{2}$As$_{2}$ could be tuned
superconducting through similar doping strategies. Indeed,
superconductivity with $T_c$ over 30 K has been observed in
(Eu,K)Fe$_{2}$As$_{2}$\cite{EuK} and
(Eu,Na)Fe$_{2}$As$_{2}$\cite{EuNa}.

Doping at the Fe site in EuFe$_{2}$As$_{2}$ takes advantage of
inducing possible superconductivity while leaving the magnetic
Eu$^{2+}$ layers intact, which could provide us insight to the
interplay between superconductivity and magnetism. Here we report a
systematic study on the physical properties in
EuFe$_{2-x}$Ni$_{x}$As$_{2}$ (0$\leq$\emph{x}$\leq$0.2) system. It
was found that both the SDW ordering of Fe moments and the AFM
ordering of Eu$^{2+}$ moments were suppressed by substituting Fe
with Ni. Ferromagnetic (FM) ordering of Eu$^{2+}$ moments emerges
for \emph{x}$\geq$0.06. While the SDW transition is completely
suppressed for \emph{x}$\geq$0.16, no superconducting transition was
observed down to 2 K in EuFe$_{2-x}$Ni$_{x}$As$_{2}$, in contrast
with the superconductivity in
BaFe$_{2-x}$Ni$_{x}$As$_{2}$\cite{Ni-doping Ba}. Our results suggest
a strong coupling between the magnetism of Eu$^{2+}$ ions and the
conduction electrons of [Fe$_{2-x}$Ni$_{x}$As$_{2}$]$^{2-}$ layers.

\section{\label{sec:level1}EXPERIMENT}

Polycrystalline samples of EuFe$_{2-x}$Ni$_{x}$As$_{2}$ ($x$= 0,
0.03, 0.06, 0.09, 0.12, 0.16 and 0.2) were synthesized by solid
state reaction with EuAs, Fe$_{2}$As and Ni$_{2}$As. EuAs was
presynthesized by reacting Eu grains and As powders in evacuated
silica tube at 873 K for 10 h then 1123 K for 36 h. Fe$_{2}$As was
presynthesized by reacting Fe powers and As powders at 873 K for 10
h and 1173 K for 2.5 h. Ni$_{2}$As was presynthesized by reacting Ni
powders and As powders at 873 K for 10 h then 1073 K for another 10
h. The powders of EuAs, Fe$_{2}$As and Ni$_{2}$As were weighed
according to the stoichiometric ratio, thoroughly ground and pressed
into pellets in an argon-filled glove-box. The pellets were sealed
in evacuated quartz tubes and annealed at 1173 K for 24 h and
furnace-cooled to room temperature.
\begin{figure}
\includegraphics[width=7.5cm]{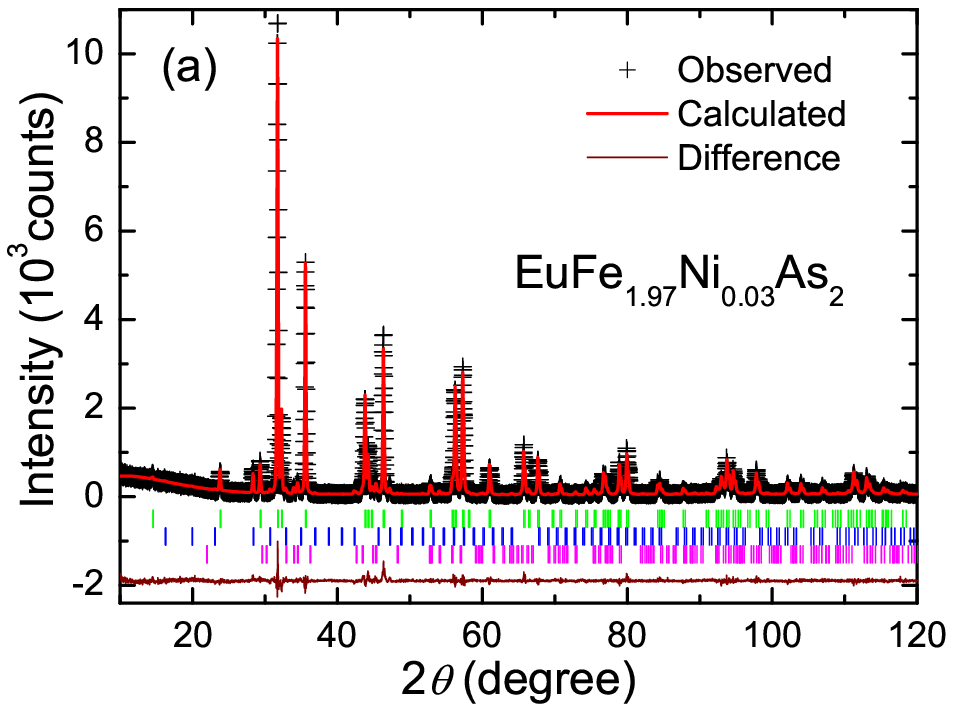}
\includegraphics[width=7.5cm]{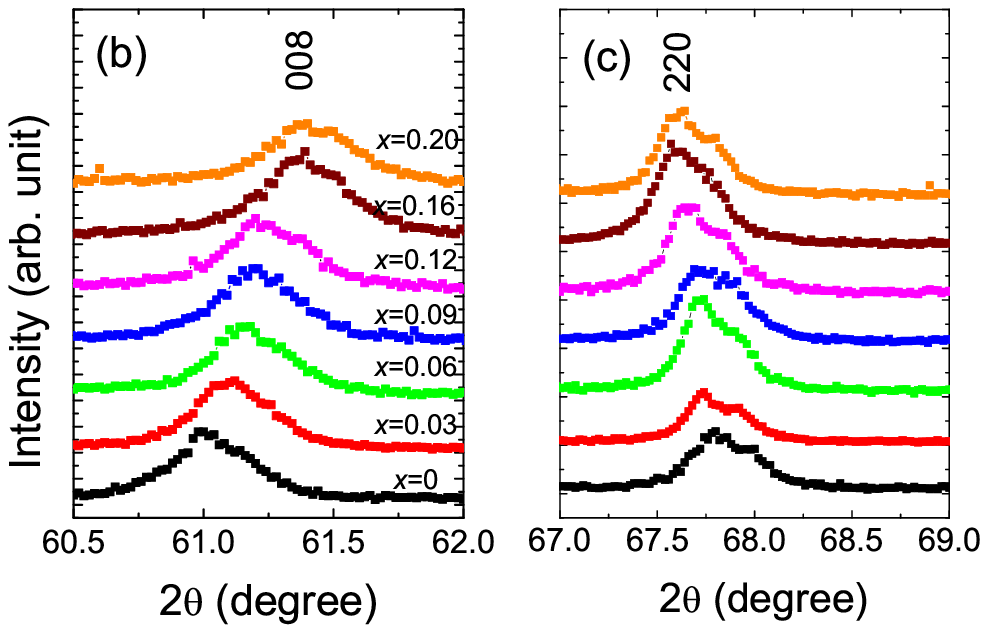}
\includegraphics[width=7.5cm]{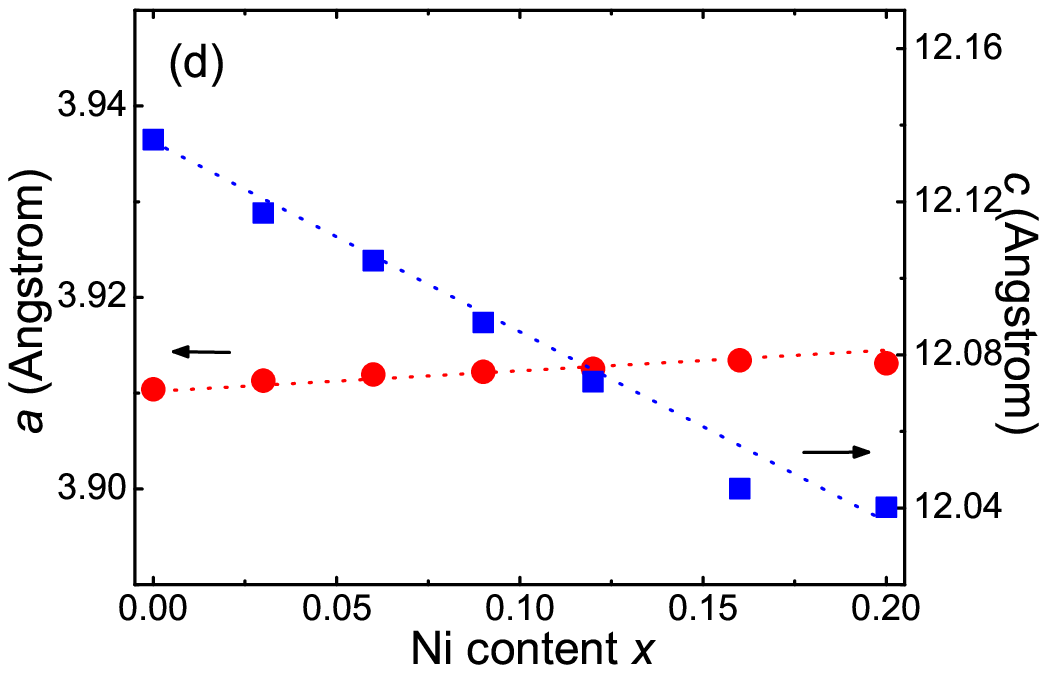}
\caption{(Color online) (a) X-ray powder diffraction pattern at room
temperature and the Rietveld refinement profile for the
EuFe$_{1.97}$Ni$_{0.03}$As$_{2}$ sample. Eu$_{2}$O$_{3}$
($\sim$1.4\%) and Fe$_{0.985}$Ni$_{0.015}$As ($\sim$6\%) are also
included in the refinement. (b) and (c) represent the (008) and
(220) diffraction peaks for the EuFe$_{2-x}$Ni$_{x}$As$_{2}$
samples, respectively. (d) Refined lattice parameters plotted as
functions of Ni content \emph{x}.}
\end{figure}
Powder x-ray diffraction (XRD) was performed at room temperature
using a D/Max-rA diffractometer with Cu-K$_{\alpha}$ radiation and a
graphite monochromator. The data were collected with a step-scan
mode. The structural refinements were performed using the programme
RIETAN 2000.\cite{Izumi} The electrical resistivity was measured
using a standard four-probe method. The measurements of dc magnetic
properties were performed on a Quantum Design Magnetic Property
Measurement System (MOMS-5). Thermopower measurements were carried
out in a cryogenic refrigerator down to 17 K by a steady-state
technique with a temperature gradient $\sim$ 1 K/cm.

\section{\label{sec:level1}RESULTS AND DISCUSSION}

The crystal structure for all the EuFe$_{2-x}$Ni$_{x}$As$_{2}$ ($x$=
0, 0.03, 0.06, 0.09, 0.12, 0.16, 0.2) samples at room temperature
were refined with the tetragonal ThCr$_{2}$Si$_{2}$ structure. An
example of the refinement profile for
EuFe$_{1.97}$Ni$_{0.03}$As$_{2}$ is shown in Fig. 1(a). The weighted
pattern factor and goodness of fit are $R_{wp}$ $\sim$ 11.2\% and
\emph{S}$\sim$1.6, indicating a fairly good refinement. Minor
impurity phases of Eu$_{2}$O$_{3}$ and Fe$_{0.985}$Ni$_{0.015}$As
are also identified. In addition, the refined occupancies are close
to the nominal value. With increasing Ni content, the (008)
diffraction peaks shift towards higher angles (Fig. 1(b)) while the
(220) diffraction peak shift towards lower angles (Fig. 1(c)). This
observation is consistent with the result from the Rietveld
refinements, which show that \emph{a}-axis increases slightly while
\emph{c}-axis shrinks remarkably with increasing Ni content, as
shown in Fig. 1(d).

\begin{figure}
\includegraphics[width=8cm]{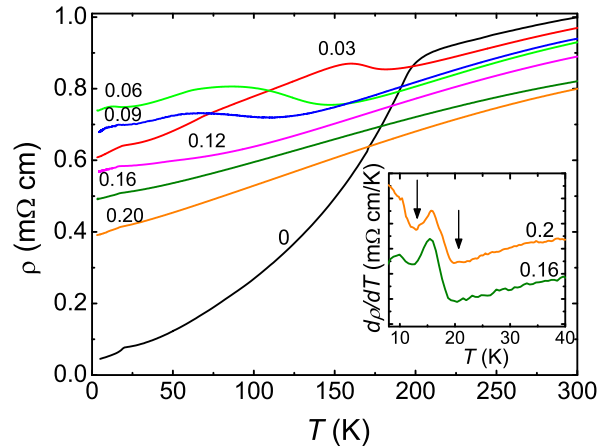}
\caption{(Color online) Temperature dependence of resistivity for
the EuFe$_{2-x}$Ni$_{x}$As$_{2}$ samples. The inset shows derivative
plots for \emph{x}=0.16 and 0.2 below 40 K. The anomalies are marked
by arrows.}
\end{figure}

Figure 2 shows the temperature dependence of normalized resistivity
($\rho$) for the EuFe$_{2-x}$Ni$_{x}$As$_{2}$ samples. The $\rho$
value at 300 K decreases with increasing Ni content, which is
probably attributed to the increase of carrier concentration induced
by the Ni doping. For the parent compound, $\rho$ drops rapidly
below 195 K and shows a kink at $\sim$20 K. The former is associated
with a SDW transition of Fe moments while the latter is due to the
AFM ordering of Eu$^{2+}$ moments.\cite{EuFeAs Ren} On Ni doping,
the anomaly in $\rho$ associated with the SDW transition is
presented as an upturn, followed by a hump. This behavior resembles
that observed in BaFe$_{2-x}$Ni$_{x}$As$_{2}$
crystals.\cite{Ni-doping Ba} With increasing Ni content \emph{x},
$T_{\text{SDW}}$ shifts to lower temperatures. For \emph{x} $\geq$
0.16 the SDW transition is completely suppressed, however, no
superconducting transition was observed down to the lowest
temperature in the present study. Instead, two kinks in $\rho$ at
low temperatures are present, which can be seen more clearly in the
derivative plots as shown in the inset of Fig. 2. It is probable
that they share the same origin as that of undoped
EuFe$_{2}$As$_{2}$ under magnetic fields, which is related to the
different magnetic states of Eu$^{2+}$ moments\cite{EuFeAs meta}.
\begin{figure}
\includegraphics[width=8cm]{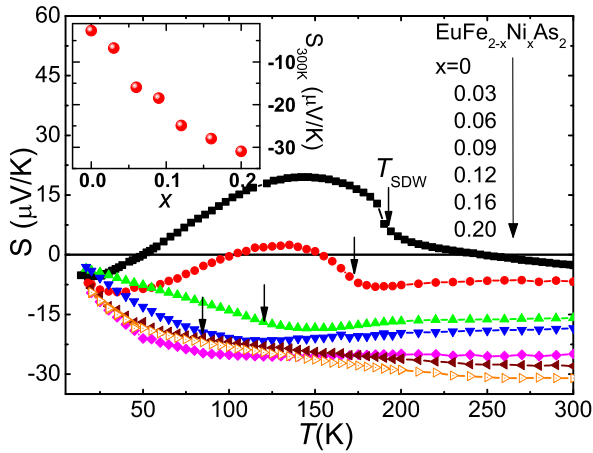}
\caption{(Color online) Temperature dependence of thermopower  for
the EuFe$_{2-x}$Ni$_{x}$As$_{2}$ samples. The inset shows the
thermopower value at 300 K plotted as a function of Ni content
\emph{x}.}
\end{figure}

Figure 3 shows the temperature dependence of thermopower ($S$) for
EuFe$_{2-x}$Ni$_{x}$As$_{2}$ samples. The sign reversal behavior,
which manifests multi-band scenario, is observed for \emph{x}=0 and
0.03. The value of $S$ for the other samples is negative. With
increasing Ni content, the room-temperature thermopower is pushed
toward more negative values, as shown in the inset of Fig. 3. For a
simple two-band model with electrons and holes, $S$ can be expressed
as,
\begin{equation}
S=\frac{n_{h}\mu_{h}|S_{h}|-n_{e}\mu_{e}|S_{e}|}{n_{h}\mu_{h}+n_{e}\mu_{e}},
\end{equation}
where $n_{h(e)}$, $\mu_{h(e)}$ and $|S_{h(e)}|$ denote the
concentration, mobility and thermopower contribution of the
holes(electrons), respectively. Therefore, the increase in $|S|$
suggests that Ni doping increases the electron concentration.
Meanwhile, the anomaly due to the SDW transition is suppressed to
lower temperatures and is no longer visible for \emph{x}=0.16, in
agreement with the above resistivity measurements. Recently, it was
found that there exists enhanced thermopower in the superconducting
window of SmFe$_{1-x}$Co$_x$AsO system.\cite{Co-doping Cao} In the
present system, no such enhancement was observed, which may be
related to the absence of superconductivity.
\begin{figure}
\includegraphics[width=7.5cm]{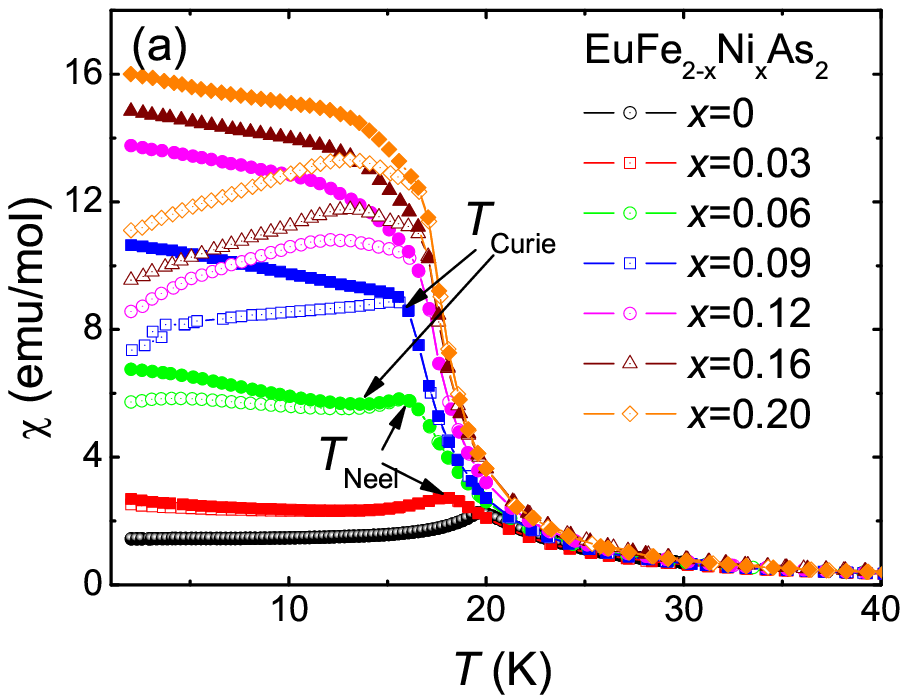}
\includegraphics[width=7.5cm]{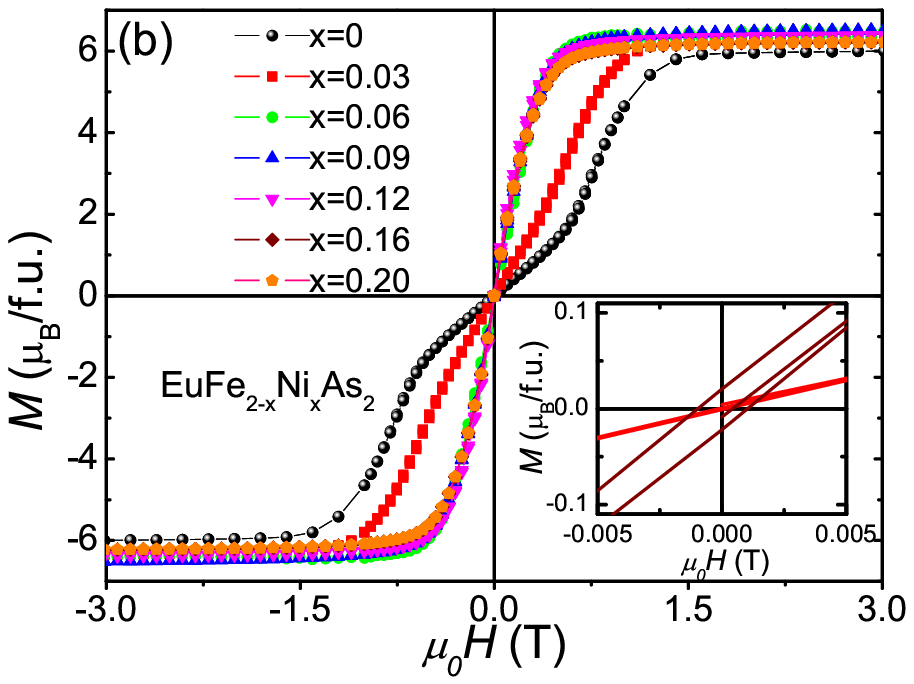}
\caption{(Color online) (a) Temperature dependence of zero-field
cooling (ZFC) (open symbols) and field-cooling (FC) (solid symbols)
magnetic susceptibility for the EuFe$_{2-x}$Ni$_{x}$As$_{2}$
samples. (b) Field dependence of magnetization at 2 K for the
EuFe$_{2-x}$Ni$_{x}$As$_{2}$ samples. The inset shows an expanded
plot of the low field region for \emph{x}=0.03 and 0.16.}
\end{figure}

Figure 4(a) shows the temperature dependence of magnetic
susceptibility ($\chi$) for the EuFe$_{2-x}$Ni$_{x}$As$_{2}$ samples
below 50 K under an applied field of 20 Oe. The $\chi$ data of 25
K$\leq$\emph{T}$\leq$180 K for \emph{x}$\geq$0.03 basically fall
onto the same curve, which can be well fitted by the modified
Curie-Weiss law,
\begin{equation}
\chi=\chi_0+\frac{C}{T-\theta},
\end{equation}
where $\chi_0$ denotes the temperature-independent term, $C$ the
Curie-Weiss constant and $\theta$ the paramagnetic Curie
temperature. The refined parameters are \emph{C}= 8.0(1)
emu$\cdot$K/mol and $\theta$=19(1) K. The calculated effective
moment $P_{eff}$ is $\sim$8 $\mu_{B}$ per formula unit, close to the
theoretical value of 7.94 $\mu_{B}$ for a free Eu$^{2+}$ ion. It is
evident that the valence state of Eu ions remains +2 and
ferromagnetic interaction between Eu$^{2+}$ moments dominates up to
10\% Ni doping. The anomaly in susceptibility due to the SDW
transition is hardly observed even after subtracting the Curie-Weiss
contribution of Eu$^{2+}$ moments. On further cooling, a sharp peak
can be observed in both $\chi_{\text{ZFC}}$ and $\chi_{\text{FC}}$
for \emph{x}=0.03 at $\sim$ 19 K, similar to that observed in
undoped EuFe$_{2}$As$_{2}$.\cite{EuFeAs Ren} We ascribe this peak to
the AFM ordering of Eu$^{2+}$ moments. With increasing Ni content to
0.06, the peak shifts to $\sim$16 K. Surprisingly, for the same
sample, a small bifurcation between ZFC and FC curves develops below
$\sim$13 K, suggesting the formation of ferromagnetic domains. For
\emph{x}$\geq$0.09, an obvious divergence between
$\chi_{\text{ZFC}}$ and $\chi_{\text{FC}}$ is seen, suggesting the
emergence of FM ordered state. It is also noted that there exists a
broad peak below $T_{\text{Curie}}$ in the ZFC curves for
\emph{x}$\geq$0.12. Interestingly, $T_{\text{Curie}}$ and
$T_{\text{Peak}}$ coincide with aforementioned two kinks in $\rho$
at low temperatures, respectively. In EuFe$_{2}$As$_{2}$ single
crystals, we have observed a metamagnetic phase with applied field
perpendicular to the $c$-axis\cite{EuFeAs meta}. Thus we speculate
that $T_{\text{Peak}}$ may be related to a successive metamagnetic
transition.

Figure 4(b) shows the field dependence of magnetization for the
EuFe$_{2-x}$Ni$_{x}$As$_{2}$ samples at 2 K. For \emph{x}=0.03, a
slope change in the \emph{M-H} curve can be seen clearly at
$\mu_{0}H$=0.55 T, which is ascribed to a field-induced metamagnetic
transition.\cite{EuFeAs Ren,EuFeAs meta,EuFeAs Chen} Moreover, there
is no hysteresis loop in the low field region, consistent with the
AFM ground state of Eu$^{2+}$ moments. For the other samples,
however, \emph{M} increases steeply with initial increasing
\emph{H}. In addition, small hysteresis loops are observed. These
results are in agreement with the above susceptibility measurements,
suggesting that Eu$^{2+}$ moments are FM ordered for
\emph{x}$\geq$0.06. It is noted that all the saturated magnetic
moments are around 6.3 $\mu_{B}$ per formula, which is smaller than
the theoretical value of 7 $\mu_{B}$ for a free Eu$^{2+}$ ion. This
discrepancy is attributed to presence of impurity phases, whose
magnetic response is much weaker.
\begin{figure}
\includegraphics[width=8.5cm]{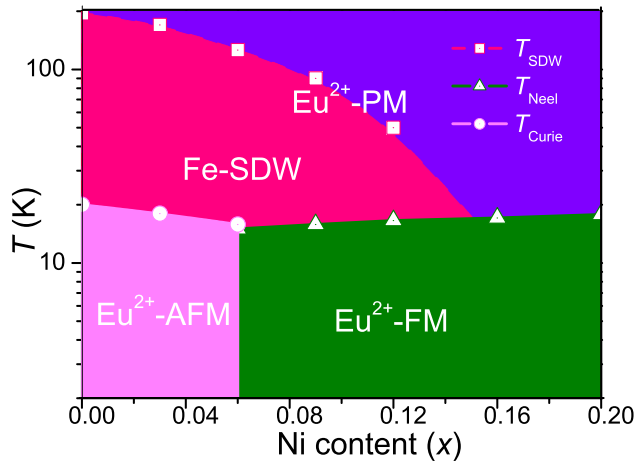}
\caption{(Color online) Magnetic phase diagram for
EuFe$_{2-x}$Ni$_{x}$As$_{2}$ system (0$\leq$\emph{x}$\leq$0.2).}
\end{figure}

Our experimental results on the physical properties of the
EuFe$_{2-x}$Ni$_{x}$As$_{2}$ system are summarized in the magnetic
phase diagram in Fig. 5. The parent compound EuFe$_{2}$As$_{2}$
shows AFM ordering of Eu$^{2+}$ moments at 20 K as well as SDW
ordering of Fe moments at 195 K. With Ni doping, both the orderings
are suppressed. On one hand, the SDW transition is gradually
suppressed and eventually disappears at $x$=0.16. Nevertheless, no
superconductivity was observed down to 2 K. On the other hand, the
magnetic ordering of Eu$^{2+}$ moments changes from AFM to FM at
$x$$\approx$0.06. This observation is surprising in view of the AFM
ordering of Eu$^{2+}$ moments in both the end members
EuFe$_{2}$As$_{2}$ and EuNi$_{2}$As$_{2}$\cite{EuFeAs Moss}. By
contrast, $T_{Neel}$ remains nearly unchanged upon 10\% Fe doping in
EuNi$_{2}$As$_{2}$.\cite{Ni122}

The AFM structure of Eu$^{2+}$ moments in EuFe$_{2}$As$_{2}$ is
proposed to be of \emph{A}-type, \emph{i.e.} FM coupling for
intralayer Eu$^{2+}$ moments while AFM coupling for interlayer
Eu$^{2+}$ moments.\cite{EuFeAs Ren,EuFeAs meta,EuFeAs Chen}. The
distance between nearest Eu$^{2+}$ layers is $\sim$ 6 {\AA} hence
direct overlap of interlayer Eu 4\emph{f} orbitals can be neglected.
Therefore, the AFM exchange between interlayer Eu$^{2+}$ moments is
probably ascribed to the carrier-mediated
Ruderman-Kittel-Kasuya-Yosida (RKKY) interaction.\cite{EuFeAs Moss}
The RKKY exchange coupling \emph{J}$_{RKKY}$ $\propto$
-$\frac{\alpha cos \alpha-sin \alpha}{\alpha^{4}}$, where
$\alpha$=2$k_{F}R$, \emph{R} denotes the distance between two
magnetic moments and $k_{F}$ the Fermi vector. One can see that
\emph{J}$_{RKKY}$ oscillates between AFM (negative) and FM
(positive) at the variation of 2$k_{F}R$. Considering the
dimensionality of the Fermi surfaces, it is probably that heavy 3-D
hole pocket derived from Fe \emph{d}$_{z}$ states\cite{Singh} is
responsible for mediating the RKKY interaction. Substitution of Fe
with Ni introduces electrons, which results in the decrease of
$k_{F}^{z}$. Meanwhile, $R$ is also shortened, as indicated by the
reduction of \emph{c}-axis. Thus the interlayer coupling may be
tuned from AFM to FM. On the other hand, the FM interaction within
the Eu$^{2+}$ layers persists up to 10\% Ni doping.  As a
consequence, a FM ordering of Eu$^{2+}$ moments is established. In
contrast, the dominant interaction between Eu$^{2+}$ moments in
EuNi$_{2}$As$_{2}$ is antiferromagnetic, as indicated by negative
paramagnetic Curie temperature\cite{EuFeAs Moss}. This may account
for the robust AFM ordering of Eu$^{2+}$ moments upon Fe doping in
EuNi$_{2}$As$_{2}$. The clarification of these issues relies on
further ARPES as well as neutron diffraction studies.

In the iron-based arsenides, superconductivity generally emerges as
the SDW order is suppressed by the carrier doping. As a matter of
fact, superconductivity with the maximum $T_{c}$ of $\sim$ 20 K has
been observed in BaFe$_{2-x}$Ni$_{x}$As$_{2}$ system.\cite{Ni-doping
Ba} Thus, the absence of superconductivity in
EuFe$_{2-x}$Ni$_{x}$As$_{2}$ may be relevant to the magnetism of
Eu$^{2+}$ ions. The RKKY interaction mentioned above may hinder the
Cooper pairing for superconductivity. Recently, re-entrant
superconducting behavior has been observed in a high pressure study
of EuFe$_{2}$As$_{2}$ crystal.\cite{EuFeAs reentrant} The results
suggest that once $T_{c}$ becomes smaller than the magnetic ordering
temperature of Eu$^{2+}$ moments, superconductivity will be
completely suppressed. If EuFe$_{2-x}$Ni$_{x}$As$_{2}$ were
superconducting, its maximum $T_{c}$ would be $\sim$ 6 K smaller
than that of BaFe$_{2-x}$Ni$_{x}$As$_{2}$ due to the existence of
paramagnetic Eu$^{2+}$ ions\cite{EuFeAs reentrant}. The assumed
$T_{c}$ is below the Curie temperatures. This could account for the
absence of superconductivity in EuFe$_{2-x}$Ni$_{x}$As$_{2}$ system.

\section{\label{sec:level1}Conclusion}

In summary, we have systematically studied the transport and
magnetic properties on a series of EuFe$_{2-x}$Ni$_{x}$As$_{2}$
polycrystalline samples with 0$\leq$\emph{x}$\leq$0.2. It is found
that both the SDW transition associated with the Fe moments and the
AFM ordering of Eu$^{2+}$ moments are suppressed upon Ni doping.
Though the SDW transition is completely suppressed for
\emph{x}$\geq$0.16, no superconducting transition is observed down
to 2 K. Surprisingly, a FM ground state of Eu$^{2+}$ moments emerges
for \emph{x}$\geq$0.06. A detailed magnetic phase diagram is
presented and discussed within the RKKY framework. Our results
suggest there exists a strong coupling the magnetism of Eu$^{2+}$
ions and the electronic state in the
[Fe$_{2-x}$Ni$_{x}$As$_{2}$]$^{2-}$ layers.

\begin{acknowledgments}
We would like to thank J. H. Dai and Q. Si for helpful discussions.
This work is supported by the National Basic Research Program of
China (No.2006CB601003 and 2007CB925001) and the PCSIRT of the
Ministry of Education of China (IRT0754).
\end{acknowledgments}


\begin{thebibliography}{00}

\bibitem{Hosono}
Y. Kamihara, T. Watanabe, M. Hirano, and H. Hosono, J. Am. Chem. Soc. \textbf{130}, 3296 (2008).
\bibitem{Chen-Sm}
X. H. Chen, T. Wu, G. Wu, R. H. Liu, H. Chen, and D. F. Fang, Nature
\textbf{453}, 761 (2008).
\bibitem{WNL-Ce}
G. F. Chen, Z. Li, D. Wu, G. Li, W. Z. Hu, J. Dong, P. Zheng, J. L.
Luo, and N. L. Wang, Phys. Rev. Lett. \textbf{100}, 247002 (2008).
\bibitem{Ren-Pr}
Z. A. Ren, J. Yang, W. Lu, W. Yi, G. C. Che, X. L. Dong, L. L. Sun,
and Z. X. Zhao, Materials Research Innovations \textbf{12}, 105
(2008).
\bibitem{Ren-Nd}
Z. A. Ren, J. Yang, W. Lu, W. Yi, X. L. Shen, Z. C. Li, G. C. Che,
X. L. Dong, L. L. Sun, F. Zhou, and Z. X. Zhao, Europhysics Lett.
\textbf{82}, 57002 (2008).
\bibitem{WHH} H. H. Wen, G. Mu, L. Fang, H. Yang, and X. Y. Zhu, Europhysics Lett. \textbf{82}, 17009 (2008).
\bibitem{Wang-Th}
C. Wang, L. J. Li, S. Chi, Z. W. Zhu, Z. Ren, Y. K. Li, Y. T. Wang,
X. Lin, Y. K. Luo, S. Jiang, X. F. Xu, G. H. Cao, and Z. A. Xu,
Europhysics Lett. \textbf{83}, 67006 (2008).
\bibitem{Johnson&Jeitschko}V. Johnson and W. Jeitschko, J. Solid State Chem. \textbf{11}, 161 (1974).
\bibitem{Quebe}P. Quebe, L. J. Terbuchte, and W. Jeitschko, Journal of Alloys and Compounds \textbf{302}, 70(2000).
\bibitem{Pfisterer1980}M. Pfisterer, and G. Nagorsen, Z. Naturforsch. B: Chem. Sci. \textbf{35}, 703 (1980).
\bibitem{Pfisterer1983}M. Pfisterer, and G. Nagorsen, Z. Naturforsch. B: Chem. Sci. \textbf{38}, 811 (1983).
\bibitem{EuFeAs} R. Marchand, W. Jeitschko, J.
Solid State Chem. \textbf{24}, 351 (1978).
\bibitem{CaFeAs ChenXH} G. Wu, H. Chen, T. Wu, Y. L. Xie, Y. J. Yan, R. H. Liu,
X. F. Wang, J. J. Ying, and X. H. Chen, J. Phys.: Condens. Matter
\textbf{20}, 422201 (2008).
\bibitem{BaFe2As2}
M. Rotter, M. Tegel, D. Johrendt, I. Schellenberg, W. Hermes, and R.
P\"{o}ttgen, Phys. Rev. B \textbf{78}, 020503(R) (2008).
\bibitem{DaiPC Neutron}
C. de la Cruz, Q. Huang, J. W. Lynn, J. Li, W. Ratcliff II, H. A.
Mook, G. F. Chen, J. L. Luo, N. L. Wang, and Pengcheng Dai, Nature
\textbf{453}, 899 (2008).
\bibitem{Kotliar}K. Haule, J. H. Shim, and G. Kotliar, Phys. Rev. Lett. \textbf{100}, 226402
(2008).
\bibitem{Cao}C. Cao, P. J. Hirschfeld, and H. P. Cheng, Phys. Rev. B \textbf{77},
220506(R) (2008).
\bibitem{Singh}D. J. Singh, and M. H. Du, Phys. Rev. Lett. \textbf{100}, 237003
(2008).

\bibitem{Co-doping Sefat} A. S. Sefat, A. Huq, M. A. McGuire, R. Jin, B. C.
Sales, D. Mandrus, L. M. D. Cranswick, P. W. Stephens, and K. H.
Stone, Phys. Rev. B \textbf{78}, 104505 (2008).
\bibitem{Co-doping Cao} C. Wang, Y. K. Li, Z. W. Zhu, S. Jiang, X. Lin, Y. K. Luo, S. Chi, L. J. Li, Z. Ren, M. He, H. Chen, Y. T. Wang, Q. Tao,
G. H. Cao, and Z. A. Xu, Phys. Rev. B \textbf{79}, 054521 (2009).
\bibitem{Ni-doping Cao} G. H. Cao, S. Jiang, X. Lin, C. Wang, Y. K.
Li, Z. Ren, Q. Tao, J. H. Dai, Z. A. Xu, and F. C. Zhang,
arXiv:0807.4328.
\bibitem{Co-doping Ba} A. S. Sefat, R. Jin, M. A. McGuire, B. C.
Sales, D. J. Singh, and D. Mandrus, Phys. Rev. Lett. \textbf{101},
117004 (2008).
\bibitem{Ni-doping Ba} L. J. Li, Y. K. Luo, Q. B. Wang, H. Chen, Z. Ren, Q.
Tao, Y. K. Li, X. Lin, M. He, Z. W. Zhu, G. H. Cao, and Z. A. Xu,
New J. Phys., \emph{in press}.
\bibitem{Co-doping SrFe2As2} A. Leithe-Jasper, W. Schnelle, C. Geibel, and H. Rosner, Phys. Rev. Lett. \textbf{101},
207004 (2008).
\bibitem{EuFeAs Ren} Z. Ren, Z. W. Zhu, S. Jiang, X. F. Xu, Q. Tao,
C. Wang, C. M. Feng, G. H. Cao, and Z. A. Xu, Phys. Rev. B
\textbf{78}, 052501 (2008).
\bibitem{EuFeAs jeevan} H. S. Jeevan, Z. Hossain, D. Kasinathan, H. Rosner, C.
Geibel, and P. Gegenwart, Phys. Rev. B, \textbf{78}, 052502 (2008).
\bibitem{BaK rotter} M. Rotter, M. Tegel, and D. Johrendt, Phys. Rev. Lett. \textbf{101},
107006 (2008).
\bibitem{SrK wang} G. F. Chen, Z. Li, G. Li, W. Z. Hu, J. Dong, X.
D. Zhang, P. Zheng, N. L. Wang, and J. L. Luo, Chin. Phys. Lett.
\textbf{25}, 3403 (2008).
\bibitem{SrK zhu} K. Sasmal, B. Lv, B. Lorenz, A. M. Guloy, F. Chen, Y.
Y. Xue, C. W. Chu, and Phys. Rev. Lett. \textbf{101}, 107007 (2008).
\bibitem{EuK} H. S. Jeevan, Z. Hossain, D. Kasinathan, H. Rosner, C.
Geibel, and P. Gegenwart, Phys. Rev. B, \textbf{78}, 092406 (2008).
\bibitem{EuNa} Y. P. Qi, Z. S. Gao, L. Wang, D. L. Wang, X. P.
Zhang, and Y. W. Ma, New J. Phys., \emph{in press}.
\bibitem{Izumi} F. Izumi, and T. Ikeda, Mater. Sci. Forum, \textbf{198}, 321 (2000)
\bibitem{EuFeAs meta} S. Jiang, Y, K, Luo, Z. Ren, Z. W. Zhu, C. Wang, X. F. Xu, Q. Tao,
G. H. Cao, and Z. A. Xu, New J. Phys., \emph{in press}.
\bibitem{EuFeAs Chen} T. Wu, G. Wu, H. Chen, Y. L. Xie, R. H. Liu,
X. F. Wang, and X. H. Chen, arXiv:0808.2247.
\bibitem{EuFeAs Moss} H. Raffius, E. M\"{o}rsen, B. D. Mosel, W.
M\"{u}ller-Warmuth, W. Jeitschko, L. Terb\"{u}chte, and T. Vomhof,
J. Phys. Chem. Solids \textbf{54}, 135 (1993).
\bibitem{Ni122} Z. Ren \emph{et al.}, unpublished results.
\bibitem{EuFeAs reentrant} C. F. Miclea, M. Nicklas, H. S. Jeevan,
D. Kasinathan, Z. Hossain, H. Rosner, P. Gegenwart, C. Geibel, and
F. Steglich, arXiv:0808.2026.
\end{thebibliography}
\end{document}